\documentclass[11pt]{JHEP3}

\usepackage{amssymb,amsmath}
\usepackage[]{graphicx}
\usepackage{epsfig}
\usepackage{amsfonts}

\usepackage{amscd}
\def\c{\cdot}
\def\Li{\text{Li}_2}

\def\be{\begin{eqnarray}}
\def\ee{\end{eqnarray}}

\def\e{\epsilon}

\def\d{\partial}

\newcommand{\idn}{{1\relax{\kern-.35em}1}}

\newcommand{\abs}[1]{\left\vert#1\right\vert}

\def\e{\epsilon}

\preprint{WIS/01/08-JAN-DPP}

\title{On collinear factorization of Wilson loops and MHV amplitudes in $\mathcal{N}=4$ SYM}
\author{Zohar Komargodski\\
Department of Particle Physics, The Weizmann Institute of Science\\
Rehovot 76100, Israel\\ \email{zkomargo@wisemail.weizmann.ac.il}}

\abstract{We consider the (multi) Splitting function of Wilson
loops and MHV gluon scattering S matrix elements in
$\mathcal{N}=4$ SYM. At strong coupling, one can utilize the
methods of Alday and Maldacena and at weak coupling (one loop) the
correspondence to light like Wilson loops is used. In both cases,
the (multi) Splitting function corresponds to flattened cusps in
the light like polygon, allowing for a clean disentanglement from
the other gluons. We compute it in some cases and estimate some
terms in other cases. We also prove the anomalous Ward identity of
Drummond et al. in the strong coupling regime. Lastly, we briefly
comment on a possible strategy for a proof of collinear
factorization of Wilson loops at higher orders of perturbation
theory.}

\begin{document}

\section{Introduction}
In this paper we analyze the collinear limit of Wilson loops, in the context of the recently conjectured
relation between gluon scattering amplitudes in $\mathcal{N}=4$ SYM and light like Wilson loops in the same
theory. This interesting relation has stemmed from an exciting result \cite{Alday:2007hr} implying such a
relation at strong coupling.\footnote{Other ideas related to \cite{Alday:2007hr} were studied in many papers
\cite{RelatedPapers}.} It is not yet clear whether this equivalence holds in perturbation theory, but it has
been proven at one loop \cite{Drummond:20074pt1loop,Brandhuber:2007yx} and for the case of four and five gluons
at two loops \cite{Drummond:20074pt2loops}\cite{Drummond:2007wardidentity}.

A prominent actor in this circle of ideas is the BDS guess \cite{Bern:2005iz} which provides an inspiring
iterative structure for MHV gluon scattering amplitudes (in $\mathcal{N}=4$ SYM), order by order in perturbation
theory. This iterative structure allows a resummation of perturbation theory and provides a natural guess for
the complete, non perturbative, gluon scattering amplitude. This ansatz  predicted correctly the strong coupling
form of four point function \cite{Alday:2007hr} (where the cusp solution of \cite{Kruczenski:2002fb} was used)
and in this paper we show explicitly that it predicts correctly the five point function at strong coupling as
well. In addition, we show that the duality to minimal surfaces in $AdS_5$ implies an anomalous Ward identity
for any $n$ (satisfied by the BDS ansatz). However, the ansatz is known to be incorrect for large enough $n$
\cite{Alday:2007he}.\footnote{It fails for large enough $n$ at strong coupling. At two loops, however, we know
that (for large enough $n$) either the ansatz or the duality to Wilson loops fails (or both).} The difference
between the BDS prediction and the correct result must be consistent with the Ward identity. This constraint is
easily satisfied by any function of conformal cross ratios.

We argue that the Splitting function of two gluons at strong coupling is consistent with the prediction of the
BDS ansatz, raising the possibility that it indeed captures correctly the Splitting function of two collinear
gluons at any 't-Hooft coupling. The case of many collinear gluons at strong coupling is analyzed using
techniques inspired by \cite{Alday:2007he}. We also discuss in detail the limit where many adjacent cusps
flatten at weak coupling. The computation of the (multi) Splitting function becomes relatively easy (at one
loop) since it requires summing a very restricted set of diagrams (In concordance with our intuition that an
almost flat cusp is a ``local" perturbation.).\footnote{The reader may wonder whether the multi Splitting
function is independent of the Splitting function of two gluons. The situation here is analogous to some well
known algebraic properties of the OPE (in a two dimensional CFT for instance). If there are some operators
distributed in a small region of space, they can be replaced by an effective single operator. However, the
structure function for this is not given by performing successive leading order OPE of pairs of operators
(unless there is a hierarchy of scales). Rather, one is forced to resum the OPE of pairs of operators. In this
paper, we consider configurations of gluons which don't have ``hierarchy in their collinearity," in fact, our
configurations are periodic. So, we expect the multi Splitting function to be independent of the Splitting
function of two gluons.} This analysis leads us to a suggestion of how this procedure is to be continued at
higher orders of perturbation theory. If correct, this implies that Wilson loops posses factorization properties
in perturbation theory (In particular, the Splitting function of two gluons is captured correctly by the BDS
ansatz.). Note that currently there is numerical evidence for this (at two loops order) in
\cite{Drummond:2007hexagon}.

This paper is organized as follows. In section 2 we review the properties of planar $\mathcal{N}=4$ SYM in the
collinear limit and the predictions for the Splitting function by \cite{Bern:2005iz}. In section 3 we discuss
the Splitting function of two collinear gluons both at one loop and at strong coupling. In section 4 we
generalize our methods from the previous section and discuss the multi Splitting function for a specific
periodic sequence of collinear gluons, again, at one loop and at strong coupling. In section 5 we present some
remarks and discuss some ideas for future work. Finally, in appendix \ref{Ward indentity} we prove an anomalous
Ward identity satisfied by gluon scattering amplitudes at strong coupling.

\section{Basics of MHV Gluon scattering amplitudes in $\mathcal{N}=4$ SYM} Denote the $L$ loop, $n$-pt color ordered partial gluon
amplitudes by $A_n^{(L)}$. We will consider only planar MHV amplitudes, where the helicity structure can be
factored out (and is the same as in the corresponding tree level amplitude). For such amplitudes we define the
reduced amplitude
$$M_n^{(L)}\equiv A_n^{(L)}/A_n^{(0)}.$$
The quantity which will keep track of the order in perturbation theory is
$$a\equiv \frac{N_c\alpha_s}{2\pi}(4\pi e^{-\gamma})^{\epsilon},$$
where $\alpha_s\equiv g_{YM}^2/4\pi$ is the 4d effective coupling constant (Note that $a$ differs from
$N_c\alpha_s/2\pi$ only by $O(\epsilon)$ terms.). This allows us to define the resummed reduced amplitude via
\begin{equation}\mathcal{M}_n=1+\sum_{L=1}^{\infty}a^LM_n^{(L)}.\end{equation}
An ansatz for this generating function was given in \cite{Bern:2005iz}. This is now known to be incomplete
\cite{Alday:2007he}, but we will see that some aspects of it are still very useful.

Lastly, the one loop amplitude $M_n^{(1)}$ is given by \cite{Bern:2005iz,Bern:1994zx} (and references therein)
$$M_n^{(1)}=-\frac1{2\epsilon^2}\sum_{i=1}^n\left(\frac{\mu^2}{-s_{i,i+1}}\right)^{\e}+F_n^{(1)}(\epsilon),$$
where $F_n^{(1)}$ is the one loop finite remainder,\footnote{If not said explicitly otherwise, we use the
signature $(+---)$.} which can be Taylor expanded in $\e$ as
$$F_n^{(1)}(\e)=F_n^{(1)}(0)+O(\e).$$
The precise form of $F_n^{(1)}(0)$ is known for any $n$ \cite{Bern:2005iz,Bern:1994zx}, and it is worth
mentioning for $n=4$ where it is particularly simple $F_4^{(1)}(0)=\frac12\ln^2(\frac{-t}{-s})+4\zeta_2$.

\subsection{Collinear limits}
The amplitudes $A_n^{(L)}$ should factorize in various channels. In the channel we discuss, nearby momenta
become collinear. For simplicity of notations we assume that two adjacent gluons become collinear, the more
general case has analogous factorization properties \cite{Kosower:1999xi}. The factorized form the amplitude has
in such a limit is \cite{Kosower:1999xi}
$$A_n^{(L)}\rightarrow \sum_{l=0}^L\sum_{\lambda=\pm}Split_{-\lambda}^{(l)}(z;k_i,\lambda_i;k_{i+1},\lambda_{i+1})A_{n-1}^{(L-l)}(..;k_P,\lambda;...),$$
where $k_P=(k_i+k_{i+1})$, $l$ counts the number of loops we used in the Splitting function, $z$ is the momentum
fraction $k_i\sim zk_P$ and $\lambda$ is the polarization. Such a factorization is somewhat analogous to the OPE
in colorless correlators. Similarly to the MHV amplitudes themselves, the helicity structure of the Splitting
functions can be swallowed in the tree level Splitting amplitude
$$Split_{-\lambda}^{(L)}(z;i,\lambda_i;i+1,\lambda_{i+1})=r_S^{(L)}(\e,z,k_P^2)Split_{-\lambda}^{(0)}(z;i,\lambda_i;i+1,\lambda_{i+1}).$$
We will denote for simplicity $s=k_P^2$.

We obtain that the general normalized amplitude $M_n^{(L)}$ behaves as follows in the collinear limit
\begin{equation}\label{collinear limit}M_n^{(L)}\rightarrow \sum_{l=0}^{L}r_S^{(l)}M_{n-1}^{(L-l)},\end{equation}
where we denote $r_S^{(0)}=1,M_n^{(0)}=1$ for convenience. It makes sense to formally resum the perturbative
Splitting amplitudes
\begin{equation}\label{Splitting amplitude}
\mathcal{R}=1+\sum_{L=1}^{\infty}a^Lr_S^{(L)}.
\end{equation}
With this, the result (\ref{collinear limit}) takes a more elegant form (we revive the explicit coupling
constant and regulator dependence)
\begin{equation}\label{factorzied collinear limit}
    \mathcal{M}_n(a;...k_i,k_{i+1}...;\epsilon,\mu)\rightarrow
    \mathcal{R}(a;z,k_P^2;\e,\mu)\mathcal{M}_{n-1}(a;...,k_P,...;\epsilon,\mu).
\end{equation}

We do not review the complete BDS guess here, but let us mention what it implies for the collinear Splitting
function. From the BDS guess it follows that the $L$ loop Splitting amplitude $r_S^{(L)}$ is related to the one
loop Splitting amplitude via the following recursive formula
\begin{equation}\label{iterative Splitting}r_S^{(L)}(\e)=X^{(L)}(r_S^{(l)}(\e))+f^{(L)}(\e)r_S^{(1)}(L\e)+O(\e),\end{equation}
where $X^{(L)}$ is a polynomial in the $r_S^{(l)}$ with $l<L$ and $f^{(L)}(\e)$ is some regular function of
$\e$. The explicit form of $X^{(L)}$ is
$$X^{(L)}(y^{(l)})=y^{(L)}-\ln(1+\sum_{j=1}^{\infty}a^jy^{(j)})\biggr |_{a^L},$$
which is a polynomial in $y^l$ with $l<L$. Plugging this into (\ref{iterative Splitting}) we get

\begin{equation}\label{exponentiated Splitting}
    \ln\left(\mathcal{R}(a;z,k_P^2;\e,\mu)\right)=\sum_{l=1}^{\infty}a^l(f^{(l)}(\e)r_S^{(1)}(l\e)+O(\e))=\sum_{l=1}^{\infty}a^lf^{(l)}(\e)r_S^{(1)}(l\e)+O(\e).
\end{equation}
The one loop Splitting amplitude is known to all orders in $\e$ \cite{Bern:1994zx}
\begin{equation}\label{one loop Splitting}r_S^{(1)}=\frac{\hat c}{\e^2}\left(\frac{\mu^2}{-s}\right)^{\e}\left[-\frac{\pi \e}{\sin(\pi\e)} \left(\frac{1-z}z\right)^{\e} +2\sum_{k=0}^{\infty}\e^{2k+1}Li_{2k+1}\left(\frac{-z}{1-z}\right) \right],\end{equation}
where $\hat c=\frac{e^{\e\gamma}}{2}\frac{\Gamma(1+\e)\Gamma^2(1-\e)}{\Gamma(1-2\e)}$ and the polylogarithms are
defined inductively $Li_n(z)=\int_0^z\frac{dt}tLi_{n-1}(t)$ with the initial condition
$Li_2(z)=-\int_0^z\frac{dt}t\ln(1-t)$. There is a subtlety we have to mention here. The total momentum of two
collinear gluons is always time like in $\mathbb{R}^{3,1}$. In order for a (real) strong coupling minimal
surface to exist \cite{Alday:2007he}, we would better work with $\mathbb{R}^{2,2}$ (In this signature two
collinear gluons can be made space like.).

\section{The Splitting function of two gluons}
\subsection{One loop aspects}

\begin{figure}\begin{center}
\epsfig{file=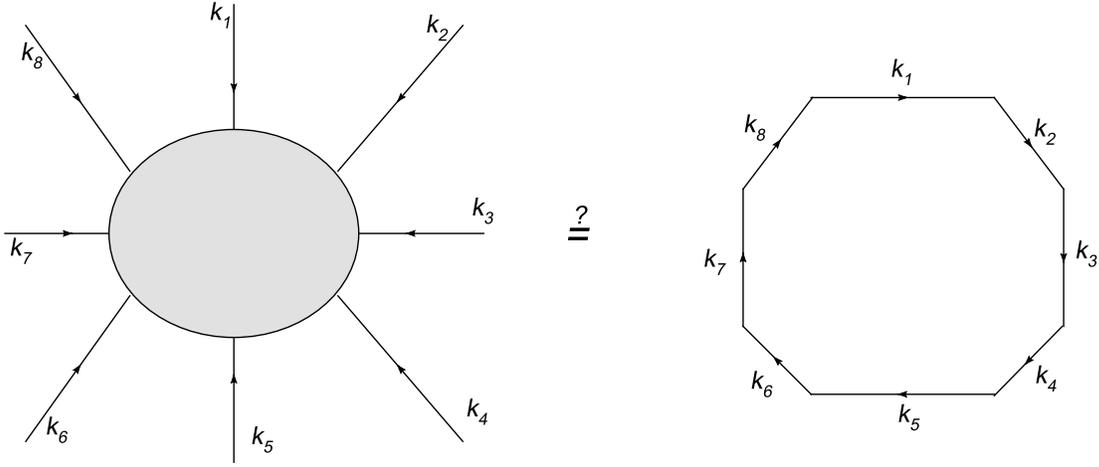} \caption{The general, conjectured, correspondence between gluon scattering
amplitudes and Wilson loops associates edges of the Wilson loop to momenta of gluons. It was shown to hold at
one loop for any number of external legs \cite{Brandhuber:2007yx}. }\label{gluonloopcorresp}.
\end{center}\end{figure}

Let us understand the Wilson loop manifestation of the collinear limit. For this sake we begin with a warm up
exercise where we compute the one loop Splitting function of two gluons from the Wilson loop representation of
one loop MHV amplitudes \cite{Drummond:20074pt1loop,Brandhuber:2007yx}. At this order, the definition of our
normalized amplitudes implies that
$$M_n^{(1)}-M_{n-1}^{(1)}\rightarrow r_S^{(1)}$$ in the collinear
limit. In \cite{Brandhuber:2007yx} it was shown that $M_{n}^{(1)}$ can be obtained from a Wilson loop
computation (at one loop), where the Wilson loop is a polygon with edges coinciding with the momenta of the
original gluons in the S matrix, see figure \ref{gluonloopcorresp}.

Physically, we expect that the only important diagrams involve edges corresponding to the collinear gluons and,
at most, their closest neighbors. We can argue that this is correct by emphasizing, first, that the Splitting
function is universal. Hence, generic diagrams connecting one of the collinear gluons with a far away line
should decouple in the collinear limit. We will show that this is indeed correct in the sequel. Concerning the
two adjacent lines to the collinear pair of gluons, it turns out that they don't decouple in a trivial manner.

\begin{figure}\begin{center}
\epsfig{file=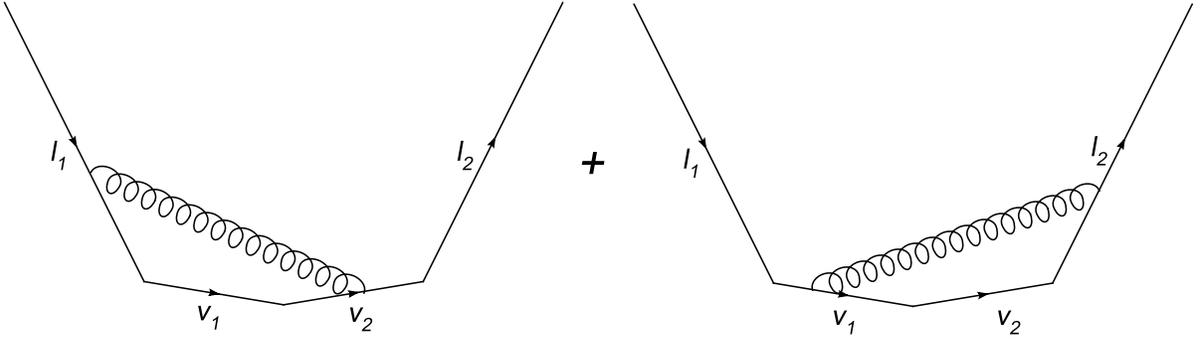} \caption{The two contributions at one loop to the difference of finite parts in
the collinear limit. }\label{curvesofWilsongluons}.
\end{center}\end{figure}

The divergent terms of gluon scattering amplitudes correspond, in the Wilson loop language (at one loop), to
propagators between two adjacent edges. In the collinear limit there are interesting contributions from small
changes in the angle of the two cusps $\{l_1,v_1\}$ and $\{v_2,l_2\}$ (see figure \ref{curvesofWilsongluons} for
the notations). This effect accounts for many of the terms in (\ref{one loop Splitting}). To evaluate these
systematically, we write first
$$r_S^{(1)}=(M_n^{(1)}-M_{n-1}^{(1)})\bigg|_{\text{collinear limit}}=(I_n^{(1)}-I_{n-1}^{(1)})\bigg|_{\text{collinear limit}}+(F_n^{(1)}-F_{n-1}^{(1)})\bigg|_{\text{collinear limit}},$$
where $I$ and $F$ are the divergent and finite parts, respectively. The contribution from
$(I_n^{(1)}-I_{n-1}^{(1)})\bigg|_{\text{collinear limit}}$ is as follows ($v_1$ and $v_2$ are collinear and
$v\equiv v_1+v_2$)
\begin{gather}\label{divcollim}
    (I_n^{(1)}-I_{n-1}^{(1)})\bigg|_{\text{collinear limit}}=\frac{-1}{2\e^2}\biggl(  (\frac{\mu^2}{-2l_1\cdot v_1 })^{\e}+(\frac{\mu^2}{-2v_1\cdot v_2})^{\e}+
    (\frac{\mu^2}{-2v_2\cdot l_2})^{\e}\cr-(\frac{\mu^2}{-2l_1\cdot v})^{\e}-(\frac{\mu^2}{-2v\cdot l_2})^{\e}\biggr)=\cr=
    \frac{-1}{2\e^2}\biggl(  (\frac{\mu^2}{-s})^{\e}-(1-z^{-\e})(\frac{\mu^2}{-2l_1\cdot v})^{\e}-
    (1-(1-z)^{-\e})(\frac{\mu^2}{-2l_2\cdot v})^{\e}\biggr)=\cr=
\frac{-1}{2\e^2}(\frac{\mu^2}{-s})^{\e}+\frac{1}{2\e}\left(\ln(z)-\frac{\e}2(\ln(z))^2\right)(\frac{\mu^2}{-2l_1\cdot
v})^{\e}+\frac{1}{2\e}\left(\ln(1-z)-\frac{\e}2(\ln(1-z))^2\right)(\frac{\mu^2}{-2l_2\cdot v})^{\e},
\end{gather}
where we have ignored $O(\e)$ terms in our final expression, as we do everywhere in the sequel.
 Note that in the first equality the result is still exact and the collinear limit is taken only in the
subsequent manipulations. The divergent terms as $\e\rightarrow 0$ in
$$(I_n^{(1)}-I_{n-1}^{(1)})\bigg|_{\text{collinear limit}},$$
agree with those in $r_S^{(1)}$ (\ref{one loop Splitting}), which is just a simple consistency check. The finite
parts should combine with the ones from
$$(F_n^{(1)}-F_{n-1}^{(1)})\bigg|_{\text{collinear limit}},$$ to
give the correct $r_S^{(1)}$. Finite parts arise in the Wilson loop computation from diagrams between non
adjacent edges. The point is that there are only two non trivial diagrams contributing the missing finite parts
of the collinear Splitting function, $\{l_1,v_2\}$ and $\{v_1,l_2\}$, depicted in figure
\ref{curvesofWilsongluons}.

The exact computation of any one loop diagram was conducted in \cite{Brandhuber:2007yx}. In our case, there is
an immediate simplification due to the fact that these are really one-mass easy box functions (the general case
includes two-mass easy box functions). Upon applying the collinear limit and dealing with the kinematics we
arrive at the following expression ($v_1\sim z(v_1+v_2)\equiv zv$) for the finite part
\begin{gather}
    -Li_2\left(1+\frac{2zv\cdot l_2}{v^2}\right)-Li_2\left(1+\frac{2(1-z)v\cdot l_1}{v^2}\right)-Li_2\left(1+\frac z{1-z}\right)-Li_2\left(1+\frac {1-z}{z}\right)+\cr+
    Li_2\left(1+\frac{z(v^2+2v\cdot l_2)}{(1-z)v^2}\right)+Li_2\left(1+\frac{(1-z)(v^2+2v\cdot
    l_1)}{zv^2}\right),
\end{gather}
where $Li_2$ is the famous Dilogarithm of Euler. This can be
further simplified in the collinear limit by noting that the
definition of the collinear limit is equivalent to $v^2\ll v\cdot
l_1$ and $v^2\ll v\cdot l_2$. Thus the contribution of the two one
loop diagrams is
\begin{gather}\label{oneloopWilson1}
    -Li_2\left(\frac{2zv\cdot l_2}{v^2}\right)-Li_2\left(\frac{2(1-z)v\cdot l_1}{v^2}\right)-Li_2\left(1+\frac z{1-z}\right)-Li_2\left(1+\frac {1-z}{z}\right)+\cr+
    Li_2\left(\frac{2zv\cdot l_2}{(1-z)v^2}\right)+Li_2\left(\frac{2(1-z)v\cdot
    l_1}{zv^2}\right).
\end{gather}
An identity by Euler
$$Li_2(z)+Li_2(1/z)+\frac12\log^2(-z)+\pi^2/6=0,$$
implies that for $|z|\gg1$
$$Li_2(z)=-\frac12\log^2(-z)-\pi^2/6+O(z).$$
Another identity we need to use is
$$Li_2(z)+Li_2(1-z)=-\log(z)\log(1-z)+\pi^2/6.$$
With these relations (\ref{oneloopWilson1}) becomes (up to an additive constant)
\begin{gather}\label{double Splitting}
2(F_n^{(1)}-F_{n-1}^{(1)})\bigg|_{\text{collinear limit}}=\cr=\frac12 \Biggl(  \ln^2\left(-\frac{-2zv\cdot
l_2}{-v^2}\right) +\ln^2\left(-\frac{-2(1-z)v\cdot l_1}{-v^2}\right) - \ln^2\left(-\frac{-2zv\cdot
l_2}{-(1-z)v^2}\right)- \ln^2\left(-\frac{-2(1-z)v\cdot
l_1}{-zv^2}\right)+\cr+\ln^2(-1/(1-z))+\ln^2(-1/z)-\ln(z)\ln(1-z) \Biggr)=\cr=
 \ln(1-z)\ln\left(\frac{-2zv\cdot l_2}{-v^2}\right)+ \ln(z)\ln\left(\frac{-2(1-z)v\cdot l_1}{-v^2}\right)-\ln(z)\ln(1-z)=\cr=
\ln(z)\ln(1-z)+\ln(1-z)\ln\left(\frac{-2v\cdot l_2}{-v^2}\right)+\ln(z)\ln\left(\frac{-2v\cdot
l_1}{-v^2}\right).
\end{gather}

One has to be a little careful in the manipulations leading to the last line, the imaginary parts cancel neatly
and the result in the end is real. Adding this result to the one for the contribution to the Splitting amplitude
from the the divergent piece, the dependence on $l_1$ and $l_2$ cancels very nicely and we are left precisely
with the expected form of the one-loop Splitting function.\footnote{Note the factor $2$ in front of
$F_n^{(1)}-F_{n-1}^{(1)}$ in (\ref{double Splitting}). }

We see that in a very geometrical manner, Wilson loops exhibit a decoupling of the collinear region from the
rest of the gluons. The non trivial part is to deal with the neighboring adjacent lines, where both the non
trivial $z$ dependence resides and where the non trivial cancelation of the $l_1,l_2$ dependence occurs. The
cancelation of $l_1,l_2$ is absolutely necessary for factorization and for the consistency of our procedure.

\subsection{Strong coupling aspects}
In appendix \ref{Ward indentity} we show that the finite part of the minimal area surface corresponding to
gluons scattering matrix elements at strong coupling has to satisfy the following anomalous Ward identity
\begin{equation}\label{diffWard1}
    \sum_i\hat{K}_i^\mu F_n=\frac{\sqrt{\lambda}}{4\pi}\sum_{i=1}^n(x^\mu_{i-1}+x^\mu_{i+1}-2x_{i}^\mu)\ln\left( s_{i,i+1}\right).
\end{equation}
In the equation above,
\begin{equation}\sum_i\hat{K}_i^\mu=\sum_i(2x_i^\mu(x_i\cdot
\d_i)-x_i^2\d_i^\mu),\end{equation} $x_{i}$ are the coordinates of the cusps and $\lambda$ is the 't Hooft
coupling constant. This equation shows explicitly that the results of \cite{Drummond:2007wardidentity} hold in
the strong coupling regime, as one may expect.\footnote{In \cite{Drummond:2007wardidentity} the same Ward
identity was proven to all loop orders using dimensionally regularized path integral. It is not clear to us
whether this technique is valid beyond perturbation theory, hence, it is useful to verify the anomalous Ward
identity directly.} This constraint fixes, up to a constant, four and five point functions. In these two cases,
the solutions of this constraint coincide with the suggestion by BDS \cite{Bern:2005iz}, again, up to a possible
overall coefficient. Thus, the Splitting function coming from the difference of a pentagon and the quadrangle
coincides with the prediction of \cite{Bern:2005iz} to the Splitting function at strong coupling (which is
roughly (\ref{one loop Splitting}) exponentiated with a pre-factor of the cusp anomalous dimension).

Since at strong coupling we are computing the minimal area of surfaces with prescribed boundary conditions, it
is reasonable to expect that a flattened cusp will modify the surface only locally. In other words, we expect a
factorization of Wilson loops at strong coupling in the collinear limit simply because the collinear limit
modifies the minimal surface only slightly (It would be nice to construct a rigorous proof of this intuitive
claim.). This argument implies that the Splitting function at strong coupling coincides with the one predicted
by BDS, although their ansatz fails in general \cite{Alday:2007he}. Qualitatively, an analogous situation is
happening at two loops \cite{Drummond:2007hexagon} where the Splitting function was observed to coincide with
the prediction of BDS but the complete ansatz fails beyond the collinear limit.\footnote{There is an important
difference between the two loop order and strong coupling. In the latter case, we know that gluon scattering
amplitudes are equivalent to Wilson loop \cite{Alday:2007hr}. In the former case it is a conjecture.}

\section{Multi Splitting function}
\subsection{One loop aspects}

\begin{figure}[tbp]
\begin{center}
\epsfig{file=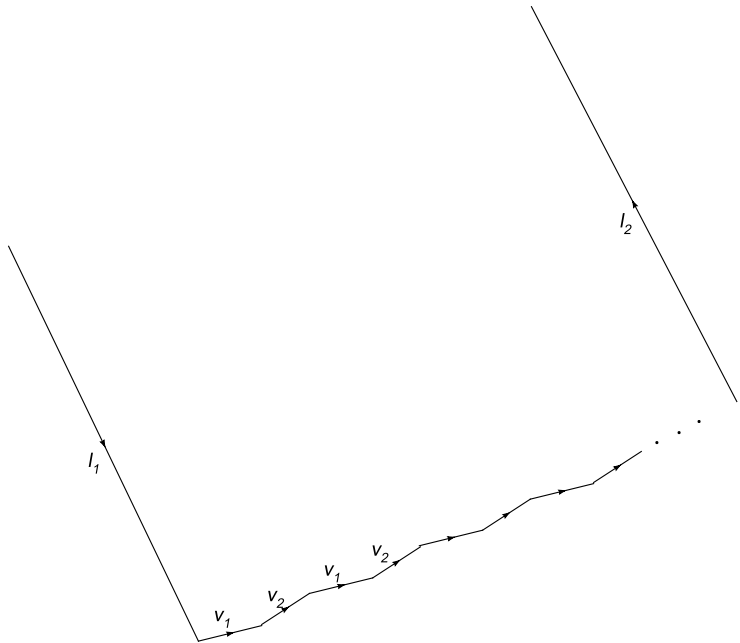,scale=0.8}\caption{The type of Wilson loop we consider. $v_1$ and $v_2$ are in
the collinear limit and the lines $l_1$ and $l_2$ are the adjacent lines to the sequence of collinear
gluons.}\label{mc}
\end{center}
\end{figure}

One may also consider $2K$ adjacent gluons with momenta $\{v_{i}| i=1...2K\}$ such that $v_i\cdot v_j$ is
smaller than all the other invariant (involving at least one gluon not from the set above) for all $i,j$.
Namely, $2K$ adjacent collinear gluons. This is possible to do, but the computation is cumbersome and the
analysis at strong coupling is less straightforward. Thus, we shall consider a special case where there are $2K$
collinear gluons whose momenta are distributed in a periodic fashion $v_1,v_2,v_1,v_2,...,v_1,v_2$. This
situation is depicted in figure \ref{mc}.

\begin{figure}[tbp]
\begin{center}
\includegraphics{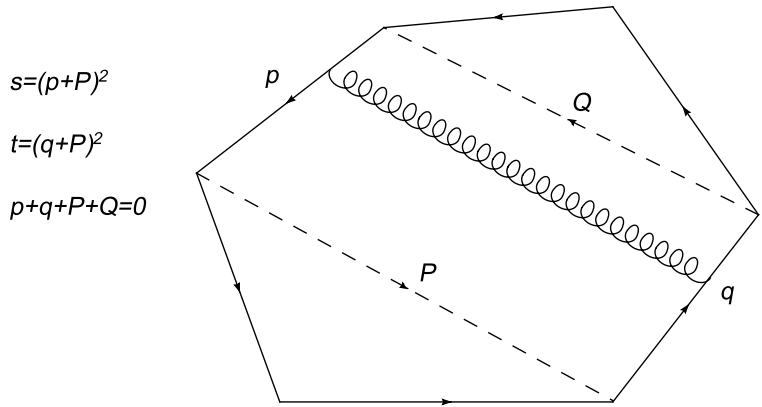}\caption{The definitions of $p,q,P,Q,s,t$ applied for a one loop contribution to a hexagon Wilson loop. Solid lines are part of the contour
defining the Wilson loop and the dashed lines just denote some effective momenta.}\label{def}
\end{center}
\end{figure}

Our first goal is to determine the multi Splitting function at one loop for such a configuration using the
technique we applied for the Splitting function of two gluons in the previous section. Again, diagrams
connecting a generic gluon with any $v_1$ or $v_2$ average out as in the Splitting function of two gluons case
and are, consequently, not important for the Splitting function. The subtlety is with the edges $l_1,l_2$ which
are crucial, for the same reasons as in the Splitting function of two gluons, in order to reproduce the correct
Splitting function. We denote $v_1\simeq z(v_1+v_2)\equiv zv$. Our notation would be that $p,q$ are the
respective momenta of edges the gluon connects to, $P$ is the overall momentum of gluons interpolating from $p$
to $q$ and $P+Q+p+q=0$. We will also use $s=(p+P)^2$, $t=(q+P)^2$. For a demonstration of these notations see
figure \ref{def}.

\begin{center}{\bf Contributions from collinear - collinear diagrams}
\end{center}
Potentially, there are contributions from diagrams where the gluon stretches along the sequence of collinear
lines. These turn out to have no non trivial kinematical information. To see it we evaluate such diagrams
explicitly. Consider, for instance, the case (where we use the notation $k\in\mathbb{N}$)
\begin{gather}
    P=kv+v_2, \hspace{1em} p=q=v_1.
\end{gather}
Obviously such diagrams vanish due to the fact they are proportional to the scalar product $v_1^2$. The less trivial case is
\begin{gather*}
    P=kv, \hspace{1em} p=v_2, \hspace{1em} q=v_1.
\end{gather*}
It turns out that the quantity defined by
$$\bar{a}\equiv\frac{P^2+Q^2-s-t}{P^2Q^2-st}$$
diverges and we can not use the formulae of \cite{Brandhuber:2007yx} to evaluate the finite part.
Instead, we solve the integral explicitly and get
\begin{gather*}
v^2\int_{[0,1]^2} d\tau_pd\tau_q\frac{1}{\left(k^2v^2+kv^2(\tau_p+\tau_q)+v^2\tau_p\tau_q\right)^{1+\e}}=\cr=
(v^2)^{-\e}\int_{[0,1]^2}
d\tau_pd\tau_q\frac{1}{\left((k+\tau_p)(k+\tau_q)\right)^{1+\e}}=(v^2)^{-\e}\left(\ln\left(\frac{k+1}{k}\right)\right)^2.\end{gather*}
Clearly, in the formal $\e$ expansion we are doing here, this contribution is a constant which we are going to
ignore anyway. We are interested only in a non trivial dependence on $z$, and these diagrams have no such.

\begin{center}{\bf Contributions from adjacent line - collinear diagrams}
\end{center}

As we have already learnt, these are the important diagrams, eventually, giving rise to a non trivial multi
Splitting function. There are actually four types of diagrams. I and II are diagrams stretching from $l_1$ to
the sequence of collinear lines (ending on $v_1$,$v_2$ respectively) and III and IV are diagrams stretching from
$l_2$ (ending on $v_2$,$v_1$ respectively). The computation of all these contributions is lengthy but
straightforward. III and IV can be inferred from I and II by simple substitutions. The result is a sum over $k$
going from $1$ to $K-1$, counting the number of $v_1,v_2$ pairs the gluon line has enclosed. In addition, two
terms corresponding to one mass easy box functions identical in form to the contributions to the Splitting
function of two gluons have to be included. We also employ the collinear limit to get rid of some terms. After
some of the dust settles down the result is
\begin{gather*}\label{I+II+III+IV}
    2(F_{n+2K}^{(1)}-F_{n+1}^{(1)})\bigg|_{\text{collinear limit}}=\cr
    \Li(\frac{v_2\c l_1}{v^2v_1\c l_1}2v\c l_1)-\Li(2\frac{v_2\c l_1}{v^2})-\Li(1+\frac{v_2\c l_1}{v_1\c l_1})+
    \Li(\frac{v_1\c l_2}{v^2v_2\c l_2}2v\c l_2)-\Li(2\frac{v_1\c l_2}{v^2})-\Li(1+\frac{v_1\c l_2}{v_2\c l_2})
     \cr
     +\sum_{k=1}^{K-1}\Biggl\{\Li(\frac{v_1\c l_1}{k^2v^2v_2\c l_1}(2kv\c l_1+2v_1\c l_1))-\Li(\frac{v_1\c l_1}{kv^2v_2\c
l_1}(2v\c l_1))-\cr-\Li(1+\frac{v_1\c l_1}{kv_2\c l_1}(k+1))+\Li(1+\frac{v_2\c l_1}{(k+1)v_1\c
l_1}k)+\Li(\frac{v_2\c l_1}{(k+1)v^2v_1\c l_1}(2v\c l_1))-\cr-\Li(\frac{v_2\c l_1}{(k+1)^2v^2v_1\c l_1}(2kv\c
l_1+2v_1\c l_1))+\Li(\frac{v_2\c l_2}{k^2v^2v_1\c l_2}(2kv\c l_2+2v_2\c l_2))-\cr-\Li(\frac{v_2\c l_2}{kv^2v_1\c
l_2}(2v\c l_2))-\Li(1+\frac{v_2\c l_2}{kv_1\c l_2}(k+1))+ \Li(1+\frac{v_1\c l_2}{(k+1)v_2\c
l_2}k)+\cr+\Li(\frac{v_1\c l_2}{(k+1)v^2v_2\c l_2}(2v\c l_2))-\Li(\frac{v_1\c l_2}{(k+1)^2v^2v_2\c l_2}(2kv\c
l_2+2v_2\c l_2))\Biggr\}.\end{gather*} This looks cumbersome but there are still many cancelations to take place
and the collinear limit was not yet fully utilized. Our next step is to turn all the dilogarithms with large
arguments into double logarithms and use the relations $v_1\sim zv$, $v_2\sim (1-z)v$.
\begin{gather}
    2(F_{n+2K}^{(1)}-F_{n+1}^{(1)})\bigg|_{\text{collinear limit}}=\cr
    \ln(z)\ln(1-z)+\ln(1-z)\ln\left(\frac{-2v\cdot l_2}{-v^2}\right)+\ln(z)\ln\left(\frac{-2v\cdot
l_1}{-v^2}\right)
     \cr
     -\frac12\sum_{k=1}^{K-1}\Biggl\{
\ln^2(\frac{-2z(k+z)v\c l_1}{(1-z)k^2v^2})-\ln^2(\frac{-2zv\c l_1}{(1-z)kv^2})+\ln^2(\frac{-2(1-z)v\c
l_1}{z(k+1)v^2})-\cr-\ln^2(\frac{-2(1-z)(k+z)v\c l_1}{z(k+1)^2v^2})+\ln^2(\frac{-2(1-z)(k+1-z)v\c
l_2}{zk^2v^2})-\ln^2(\frac{-2(1-z)v\c l_2}{zkv^2})+\cr+\ln^2(\frac{-2zv\c
l_2}{(1-z)(k+1)v^2})-\ln^2(\frac{-2z(k+1-z)v\c l_2}{(1-z)(k+1)^2v^2})\Biggr\}\cr
+\sum_{k=1}^{K-1}\Biggl\{\Li(1+\frac{(1-z)k}{z(k+1)})-\Li(1+\frac{(1-z)(k+1)}{z k})+ \Li(1+\frac{z
k}{(1-z)(k+1)})-\Li(1+\frac{z(k+1)}{(1-z)k})\Biggr\}.
\end{gather}
To decipher the physical meaning of this result, we first collect the $l_1,l_2$ dependence which gives our
expression, after some algebra, the following form
\begin{gather}\label{I+II+III+IVsimplified}
    2(F_{n+2K}^{(1)}-F_{n+1}^{(1)})\bigg|_{\text{collinear limit}}=
    \ln\left(\frac{1-z}{K}\right)\ln\left(\frac{-2v\cdot l_2}{-v^2}\right)+\ln\left(\frac{z}{K}\right)\ln\left(\frac{-2v\cdot
l_1}{-v^2}\right)+\Psi(z,K),
\end{gather}
where the function $\Psi(z,K)$ is independent, by definition, of $l_1,l_2$. We can simplify it a little and
write it as (We disregard $z$ independent constants, even if they depend on $K$. One could easily keep track of
them if needed.)
\begin{gather}\label{Psi}
    \Psi(z,K)=\cr=\ln(z)\ln(1-z)
     -2\sum_{k=1}^{K-1}\Biggl\{\ln(\frac{z}{1-z})\ln(\frac{k+z}{k+1-z})+\ln((k+1-z)(k+z))\ln(\frac{k+1}{k})\Biggr\}\cr
+\Re\sum_{k=1}^{K-1}\Biggl\{\Li(1+\frac{(1-z)k}{z(k+1)})-\Li(1+\frac{(1-z)(k+1)}{z k})+ \Li(1+\frac{z
k}{(1-z)(k+1)})-\Li(1+\frac{z(k+1)}{(1-z)k})\Biggr\}.
\end{gather}
In order to derive the equation above, we had to make sure all the imaginary contributions neatly cancel, as in
the case of the Splitting function of two gluons (Note we take the real part of the second sum in (\ref{Psi}).).

Inspecting the dependence on $l_1,l_2$ in equation (\ref{I+II+III+IVsimplified}) we see that this is exactly
what we expect to get in order for the complete Splitting function to be universal. Namely, upon adding to
(\ref{I+II+III+IVsimplified}) the contribution of the (difference of) divergent parts, we obtain an expression
independent of $l_1,l_2$. To see how it comes about, recall again that the divergent part at one loop takes the
form
\begin{gather}
    \mathcal{M}_n^{(1)}|_{div.}=\frac{-1}{2\e^2}\sum_{i=1}^n\left(\frac{\mu^2}{-s_{i,i+1}}\right)^{\e}.
\end{gather}
In our case, of an almost light like vector $Kv$, $\mathcal{M}_n^{(1)}|_{div.}-\mathcal{M}_{n-1}^{(1)}|_{div.}$
takes the form
\begin{gather}
    \frac{-1}{2\e^2}\left(\frac{\mu^2}{-2v\c l_1}\right)^{\e}\left(z^{-\e}-K^{-\e}\right)+
    \frac{-1}{2\e^2}\left(\frac{\mu^2}{-2v\c
    l_2}\right)^{\e}\left((1-z)^{-\e}-K^{-\e}\right)-\frac{2K-1}{2\e^2}\left(\frac{\mu^2}{-v^2}\right)^{\e}.
\end{gather}
Upon expanding it in $\e$, we discover that combined with (\ref{I+II+III+IVsimplified}) the dependence on the
momenta $l_1$ and $l_2$ cancels and we remain with the expression for the multi Splitting function of this
periodic sequence of collinear gluons.

We turn to investigating some properties of $\Psi(z,K)$. A trivial consistency check is that
$\Psi(z,K)=\Psi(1-z,K)$. A more interesting exercise is to check some aspects of the large $K$ behavior
dependence of $\Psi(z,K)$. The objective is to compare this result to the strong coupling behavior. To
accomplish this, we study more carefully the summands defining $\Psi(z,K)$ in (\ref{Psi})
\begin{gather}\label{eq2}
    \Re\biggl\{-2\biggl(\ln\left(\frac{z}{1-z}\right)\ln\left(\frac{1+z/k}{1+(1-z)/k}\right)+\ln\left((1+z/k)(1+(1-z)/k)\right)\ln\left
    (1+1/k\right)\biggr)
-\cr-\Li\left(1+\frac{z(1+1/k)}{(1-z)}\right)+\Li\left(1+\frac{(1-z)}{z(1+1/k)}\right)-\Li\left(1+\frac{(1-z)(1+1/k)}{
z}\right)+\cr+\Li\left(1+\frac{z }{(1-z)(1+1/k)}\right)\biggr\}.
\end{gather}
We have rewritten it in a way which enables to extract the large $k$ behavior easily. It is easy to see that the
zeroth order in the $1/k$ expansion vanishes. Thus, there is no term linear in $K$ in $\Psi(z,K)$. It is less
trivial but also straightforward (using $\Li(x)'=-\ln(1-x)/x$) to see that next term in the expansion,
proportional to $1/k$, cancels as well. Consequently, there is no term going like $\ln(K)$ in $\Psi(z,K)$.

The next, $1/k^2$ term, does not vanish but contains no $z$ dependence, so it is just a constant contribution.
The first non trivial term comes from the expansion to third order in $1/k$. This gives rise to a behaviour at
large k of the form \begin{equation}\label{psi final}\Psi(z,K)=\ln(z)\ln(1-z)-\frac23\sum_{k}^K
\left(\frac1{k^3}z(1-z)+O(\frac1{k^4})\right),\end{equation} where again some constant contribution from the
third order was thrown away.  This is convergent for large $K$, of course.

We conclude that the only possible dependence on $K$ at large K is $O(1)$. In particular, there is no non
trivial dependence on $z$ scaling with $K$. This feature can be easily compared to the Splitting function at
strong coupling.

We wish to make another important comment regarding (\ref{psi final}). Clearly, it is different from the result
one would have got had the collinear limit been taken successively (in which case there would have been a
dependence on $z$ scaling with $K$). This is indeed the expected behavior because treating successively pairs of
adjacent gluons inside the sequence is not a reliable expansion; one has to resum the expansion in $v^2$ and get
back the complete propagator.

\subsection{Strong coupling aspects}

\begin{figure}[tbp]
\begin{center}
\includegraphics{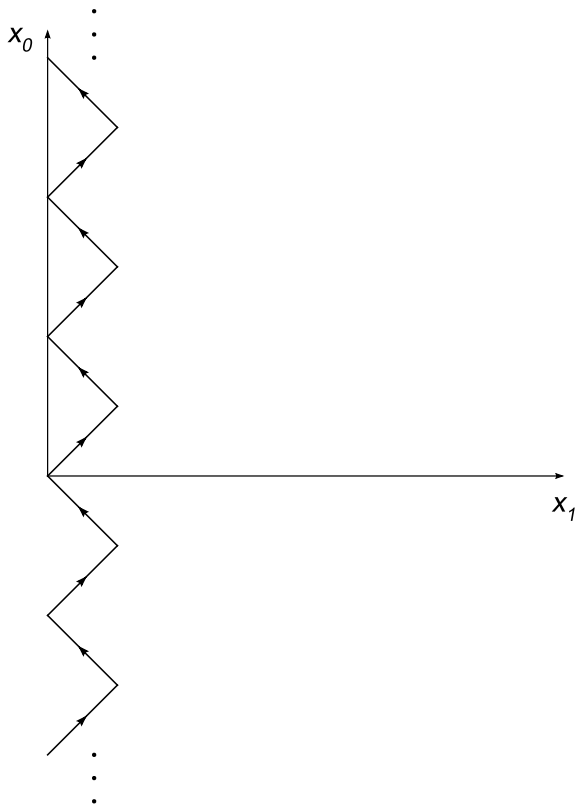}\caption{We have to look for a minimal surface ending (on the boundary of $AdS_5$) on such a periodic piecewise linear function.}\label{boundary}
\end{center}
\end{figure}

Let us first begin by studying the minimal surface whose boundary at infinity is given by a periodic sequence of
$v_1,v_2$ where $v_1$ and $v_2$ are two light like vector whose sum is time like. For definiteness we specify
$$ds^2=dx^+dx^--\sum_{i=2}(dx^i)^2,$$
with $x^{\pm}=x^0\pm x^1$. We can always Lorentz transform $v_1$ to the form $(1,0,0...,0)$. Using Euclidean
rotations in the transverse space we can bring $v_2$ to the subspace spanned by $x^+,x^-,x^2$. At this stage we
have to investigate Lorentz transformations of $\mathbb{R}^{2,1}$ which leave the line $x^+$ invariant (but can
re-scale it). There are two generators which have this property and using them we can transform any pair of
vectors to a canonical one
$$\{(1,0,0), (\alpha, \beta, \sqrt{\alpha\beta} )\}\ \rightarrow \ \{\sqrt{\beta}(1,0,0), \sqrt{\beta}(0,1,0)\}.$$
By further using the rescaling generator it follows that any pair of light like vectors whose sum is time like
can be transformed to
$$ \{(1,0,0), (0,1,0)\}.$$
Thus, it is sufficient to find the minimal surface ending on the sequence of vectors generated by $\{(1,0,0),
(0,1,0)\}$, as depicted in figure \ref{boundary}. A closely related problem was considered in
\cite{Alday:2007he}. The only (inconsequential) difference is that their the surface is rotated by $90$ degrees,
i.e the sum of the vectors is space like.

Reading out some gross features of the solution of \cite{Alday:2007he}, we see that $x^1$ decays exponentially
with the $AdS$ radius $r$ as (we use the Poincar\'e patch metric $ds^2\sim\frac{dx^2+dr^2}{r^2}$)
$$x^1(r)\sim e^{-r},$$
and because of scale and lorentz invariance, the zig-zag decays exponentially also for the original sequence
$v_1,v_2$ in the following way \begin{equation}\label{decayrate}e^{-\frac{r}{\sqrt{\abs{v_1\c
v_2}}}}.\end{equation} The area is also easy to compute, up to an unknown constant (of order one) $C$,
\begin{equation}\label{area}\mathcal{A}\sim N\frac{\mu^\e}{(v_1\c
v_2)^\e}\left(\frac1{\e^2}+\frac{1-\ln2}{2\e}+C\right)+O(\e),\end{equation} where $N=2K-1$ is just the number of
cusps (which is infinite for a strictly infinite sequence).

If such a long sequence (large K) is made collinear and embedded in some Wilson loop (part of such a Wilson loop
is depicted in figure \ref{mc}), then the relative smallness of $v_1\c v_2$ insures that very quickly the
zig-zag ``defects'' decay and one can replace the collinear lines by an effective single line, namely, there is
factorization. It is interesting to ask what is the Splitting function. There are two effects contributing to
it. The first just comes from (\ref{area}), which indeed supplies terms we expect in the Splitting function
since these are the correct divergences associated with adjacent gluons. The other kind of effects contributing
to the Splitting function are, morally, edge effects. These are out of control (as they were in
\cite{Alday:2007he}), but we can clearly expect them not to contain any non-trivial $z$ dependence scaling with
$K$. Hence, the Splitting function contains terms scaling with $K$ (exactly the ones we expect) and some non
trivial terms remaining finite as $K\rightarrow \infty$ which we can not compute.\footnote{To avoid confusion we
emphasize that in our limit $K\rightarrow\infty$  but the momentum of each individual gluon, $v_i$, remains
fixed. Namely, the total momentum of all the collinear gluons scales linearly with $K$.} This is the same
situation as for the multi Splitting function at one loop, where the non trivial information is in the function
$\Psi(z,K)$, which does not have (non trivial) terms scaling with $K$.

It will be interesting to understand better these edge effects and to compare the multi Splitting functions at
strong coupling to those predicted by the BDS ansatz, this is left to future work.

\section{More comments and possible generalization}
A crucial point at one loop, manifest in the Feynman gauge we used, is that propagators from the collinear
gluons to generic distant edges do not contribute to the Splitting function. The two adjacent edges, as we have
shown above, are an exception. One can understand this qualitatively  by noting that propagators from generic
distant edges to the collinear ones are finite as $v^2\rightarrow 0$. Hence, we expect that the difference
between a completely flattened cusp (straight line) and one which is almost flat goes to zero as $v^2\rightarrow
0$. This is indeed what happens at one loop. The two adjacent edges are exceptional because their contribution
is not finite as $v^2\rightarrow 0$ due to the integration over the region where the propagator is light like.
This is explicit, for instance, in equation (\ref{double Splitting}). Therefore, it permits a non trivial
dependence on $z$, as indeed happens.

\begin{figure}\begin{center}
\epsfig{file=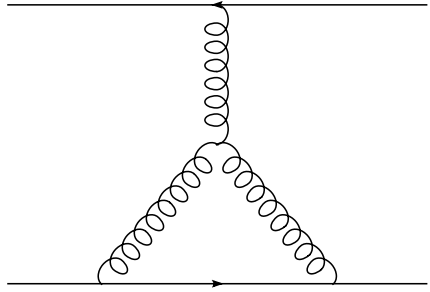} \caption{Such a diagram diverges. Both straight lines in the pictures are
light like.}\label{3vertex}.
\end{center}\end{figure}

It is not completely trivial to generalize this argument to two loops. The basic reason is that diagrams like
figure \ref{3vertex} diverge. Hence, any such diagram where two of the gluons end on two adjacent collinear
edges is expected to be large as $v^2\rightarrow 0$. At first sight, this ruins the argument completely since we
can no longer examine a very restricted set of graphs to check whether there is factorization.

There is a well known way to correct this pathology.\footnote{I am grateful to J.~Maldacena for an interesting discussion.} We can consider, instead, the axial gauge
$$n\cdot A=0,$$
with $n^2\neq 0$. It is easy to check that in this gauge the set of divergent diagrams is much simpler; they are
all ``localized" around cusp points \cite{Ellis:1978ty}, see also \cite{Drummond:2007wardidentity}. In this
gauge it should be easier to prove, to all orders, that there is factorization. We expect that this is indeed
the case (There is already numerical evidence \cite{Drummond:2007hexagon} for factorization of hexagons at two
loops.). It may be easy to show that generic edges decouple in the collinear limit, by the same token as at one
loop. The decoupling of adjacent edges is probably more tricky and we leave this study to future work.

\section*{Acknowledgments}

I would like to thank S.~Razamat for collaboration in the early stages of this project and for sharing some of
his insights. I am also grateful to O.~Aharony, M.~Berkooz, J.~Maldacena and A.~Schwimmer for useful discussions
and comments on the manuscript. This work was supported in part by the Israel-U.S. Binational Science
Foundation, by a center of excellence supported by the Israel Science Foundation (grant number 1468/06), by a
grant (DIP H52) of the German Israel Project Cooperation, by the European network MRTN-CT-2004-512194, and by a
grant from G.I.F., the German-Israeli Foundation for Scientific Research and Development.

\appendix
\section{An anomalous Ward identity at strong coupling}\label{Ward indentity}
Let us define $AdS_5$ by the embedding coordinates
$$-X_{-1}^2-X_0^2+X_1^2+X_2^2+X_3^2+X_4^2=-1,$$
and we denote
$$X_+=X_{-1}+X_4, \hspace{2em} X_-=X_{-1}-X_4. $$
We also introduce the Poincar\'e coordinates
$$X^\mu=\frac {x^\mu}{r} \ , \hspace{1em} X_+=\frac1r\ ,\hspace{1em} X_-=\frac{r^2+x^\mu x_\mu}{r}.$$
The signature used in this appendix is $(-,+,+,+)$ and $\mu=0,...,3$. The metric in Poincar\'e coordinates is
$$ds^2=\frac{dx^2+dr^2}{r^2},$$
where we set the overall scale to $1$. Inversion symmetry acts in these embedding coordinates as
$X_+\leftrightarrow X_-$ and becomes the usual inversion in $4$ dimensions ($x^\mu\rightarrow x^\mu/x^2$) near
the boundary of $AdS_5$. Special conformal transformations are realized as a combination of inversion
translation and another inversion. A simple calculation gives the precise formulas for the  action of a special
conformal symmetry on the Poincar\'e patch coordinates

\begin{gather}\label{SCT}
    r'=\frac{r}{1+2x\cdot\beta+\beta^2(r^2+x^2)},\cr
    (x')^\mu=\frac{x^\mu+(r^2+x^2)\beta^\mu}{1+2x\cdot\beta+\beta^2(r^2+x^2)}.
\end{gather}
Consider a minimal area problem whose Dirichlet boundary condition at the boundary of the Poincar\'e patch is
made of $n$ light like lines (with {\it out-in-out-in...} ordering)
\begin{gather}
    S=\frac{\sqrt{\lambda}}{2\pi}\int du_1du_2\sqrt{det \left(\d_a X^\mu\d_b X^\nu G_{\mu\nu}\right)}.
\end{gather}
This action is obviously invariant under any coordinate transformation which leaves the Poincar\'e metric
invariant, in particular (\ref{SCT}). Note that these are global transformations of the Nambu-Goto action, and
leave $u_{1}$,$u_2$ intact.

However, there are divergences in the classical area problem, and the action is supposed to be regulated. In our context,
one implements the gravitational version of Dimensional Regularization
\cite{Alday:2007hr}, where the Nambu-Goto action is slightly modified by changing the metric to
\begin{gather}\label{regmetric}
    ds^2=\frac{\sqrt{\lambda_Dc_D}}{r^\e}(\frac{dy^2+dr^2}{r^2}),\end{gather}
with
$$\lambda_D=\frac{\lambda\mu^{2\e}}{(4\pi e^{-\gamma})^{\e}}\ , \hspace{2em}c_D=2^{4\e}\pi^{3\e}\Gamma(2+\e).$$
Consequently, the Nambu-Goto action becomes
\begin{gather}
    S=\frac{\sqrt{\lambda_Dc_D}}{2\pi}\int du_1du_2\frac{1}{r^\e}\sqrt{det \left(\d_a X^\mu\d_b X^\nu
    G_{\mu\nu}\right)}\equiv \frac{\sqrt{\lambda_Dc_D}}{2\pi}\int du_1du_2\frac{\mathcal{L}_{\e=0}}{r^\e},
\end{gather}
where the $G_{\mu\nu}$ inside the square root is the same as in the unregulated metric (namely that of a
Poincar\'e patch). In order for this regulated action to be convergent, we need to take $\e<0$.

Let us perform a special conformal transformation of the fields in the action. This takes the form (\ref{SCT})
and leaves $\mathcal{L}_{\e=0}$ invariant by construction. However, the regularized action changes and becomes
\begin{gather}\label{corrected action}
    S'=\frac{\sqrt{\lambda_Dc_D}}{2\pi}\int
    du_1du_2\frac{\mathcal{L}_{\e=0}}{r^\e}\left(1+2x\cdot\beta+\beta^2(r^2+x^2)\right)^{\e}.
\end{gather}

One may worry that perhaps it is necessary to take into account corrections to the solution of the equations of
motion (since it is only a solution as long as $\e=0$). The argument here goes in the same way as in
\cite{Alday:2007hr}, where it was explained that (to reproduce the finite parts) it is sufficient to make sure
that it is $\e$ corrected (to leading order) in the vicinity of cusps. We will be careful about this rather
subtle point.

Let us study the modified action (\ref{corrected action}) and expand it in both $\e$ and $\beta$ to learn how a
small special conformal transformation modifies the regularized area of the solution. We begin by expanding in
$\e$
\begin{gather}
    S'=\frac{\sqrt{\lambda_Dc_D}}{2\pi}\int
    du_1du_2\frac{\mathcal{L}_{\e=0}}{r^\e}\times\cr\times\left(1+\e\ln\left(1+2x\cdot\beta+\beta^2(r^2+x^2)\right)+
\frac12\e^2\ln^2\left(1+2x\cdot\beta+\beta^2(r^2+x^2)\right)+O(\e^3)\right).
\end{gather}
We know that further corrections are unimportant since they are $O(\e)$ in the final answer. Expanding in
$\beta$ we see that the $\e^2$ term in the expansion is also proportional to $\beta^2$ (which means it does not
contribute to the anomalous Ward identity governing the variation under infinitesimal special conformal
transformations). Hence, the relevant terms in $S'$ are really
\begin{gather}
    S'=\frac{\sqrt{\lambda_Dc_D}}{2\pi}\int
    du_1du_2\frac{\mathcal{L}_{\e=0}}{r^\e}\times\cr\times
    \left(1+\e\ln\left(1+2x\cdot\beta+\beta^2(r^2+x^2)\right)\right)
     +O(\beta^2).
\end{gather}
Expanding the remaining logarithm to leading power in $\beta$ we get
\begin{gather}
    S'=\frac{\sqrt{\lambda_Dc_D}}{2\pi}\int
    du_1du_2\frac{\mathcal{L}_{\e=0}}{r^\e}\left(1+2\e x\cdot\beta\right)
     +O(\beta^2)=\cr=S+2\e\beta_\mu\frac{\sqrt{\lambda_Dc_D}}{2\pi}\int
    du_1du_2\frac{\mathcal{L}_{\e=0}}{r^\e}x^\mu
     +O(\beta^2).
\end{gather}

Thus, we can write our result so far as (we don't mention the $O(\beta^2)$ corrections in the sequel)
\begin{equation}\delta_{\beta} S=2\e\beta_\mu\frac{\sqrt{\lambda_Dc_D}}{2\pi}\int
    du_1du_2\frac{\mathcal{L}_{\e=0}}{r^\e}x^\mu.\end{equation}

Let us figure out how to calculate the interesting terms from this integral. Note that  $x^\mu$ is generically
finite all over the solution, both in the bulk and boundary. Since $\delta_\beta S$ contains $\e$ in front of
the integral, there are no contributions form the bulk of the worldsheet (since they are $O(\e)$). Consequently,
there are contributions only from lines and cusps. Both of these are well understood in general since they
contribute to the universal divergent parts of amplitudes.

\begin{figure}\begin{center}
\epsfig{file=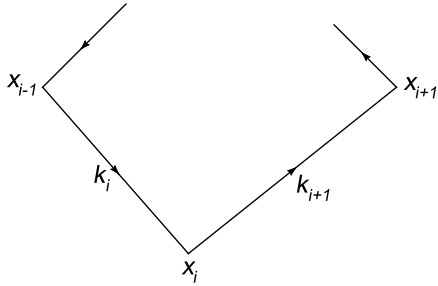}\caption{Generic cusp at the boundary of some minimal surface.} \label{cuspfig}.
\end{center}\end{figure}

Consider a cusp with the adjacent momenta being $k_i,k_{i+1}$. The cusp is assumed to be at $x_i$ and the
momentum $k_i$ begins at $x_{i-1}$ and $k_{i+1}$ ends at $x_{i+1}$ (see figure \ref{cuspfig}). We can always
make rescalings and Lorentz transformations so to bring the cusp to some plane $Y_+,Y_-$ where the boundary of
the worldsheet spans $Y_{\pm}\in[0,1]$. $Y_-=0,Y_+=1$ maps to the beginning of the vector $k_i$ in the original
frame and similarly $Y_-=1, Y_+=0$ maps to the endpoint of $k_{i+1}$. The integrand in the neighborhood of a
single cusp was analyzed in \cite{Alday:2007hr,Buchbinder:2007hm}. We may take advantage of these
results\footnote{We wish to point out a subtle point here. The ansatz of~\cite{Buchbinder:2007hm} for the
solution near cusps and lines gives correctly the divergent parts of any n-point function, but it is not clear
whether it really describes the form of the minimal surface near lines. Nevertheless, we assume that we may use
this proposal for the approximate form of the solution.} and write the contribution to the variation from a
single cusp in our case
\begin{equation}\label{cuspcontribution}
    \delta_\beta S_{i,i+1}=-2\e\beta_\mu C(\e)\frac{\sqrt{\lambda}}{8\pi}\sqrt{\frac{\mu^{2\e}}{(s_{i,i+1})^\e}}\int
    dY_+dY_-\frac{x_{(i)}^\mu(Y_+,Y_-)}{(Y_+Y_-)^{1+\e/2}},
\end{equation}
where all we need to know about $C(\e)$ is that
$$C(\e)=1+\frac{1}2(1-\ln 2)\e+O(\e^2),$$
and $s_{i,i+1}=(k_i+k_{i+1})^2=(x_{i+1}-x_{i-1})^2$. It is important to emphasize that (\ref{cuspcontribution})
includes the needed $\e$ adjustments of the single cusp solution, so we don't have to worry about this issue
anymore. In the vicinity of cusps and lines $x^\mu$ follows its Dirichlet boundary condition (which is a
piecewise linear function). Thus, we can write without loss of generality that for our single cusp contribution
    $$x_{(i)}^\mu(Y_+,Y_-)=x^\mu_{i}+\eta_{(i)} Y_++\theta_{(i)} Y_-, $$
where $\eta$ and $\theta$ are some real numbers. The integral is easily solvable
\begin{gather}\label{integral}
    \int dY_+dY_-\frac{x^\mu_{i}+\eta_{(i)} Y_++\theta_{(i)}
    Y_-}{(Y_+Y_-)^{1+\e/2}}=\cr=x^\mu_{i}\int dY_+dY_-\frac{1}{(Y_+Y_-)^{1+\e/2}}+\eta_{(i)}\int dY_+dY_-\frac{ (Y_+)^{-\e/2}}{(Y_-)^{1+\e/2}}+\theta_{(i)}
\int
dY_+dY_-\frac{(Y_-)^{-\e/2}}{(Y_+)^{1+\e/2}}=\cr=\frac{4x_{i}^\mu}{\e^2}-\frac{4(\eta_{(i)}+\theta_{(i)})}{\e(2-\e)}.
\end{gather}
We combine this factor with the rest needed for $\delta_\beta S_{i,i+1}$ to obtain
\begin{gather}
\delta_\beta S_{i,i+1}=-2\e\beta_\mu
C(\e)\frac{\sqrt{\lambda}}{8\pi}\sqrt{\frac{\mu^{2\e}}{(s_{i,i+1})^\e}}\left(\frac{4x_{i}^\mu}{\e^2}-\frac{4(\eta_{(i)}+\theta_{(i)})}{\e(2-\e)}\right)=\cr=
-2\beta_\mu
C(\e)\frac{\sqrt{\lambda}}{2\pi}\sqrt{\frac{\mu^{2\e}}{(s_{i,i+1})^\e}}\left(\frac{x_{i}^\mu}{\e}-\frac{\eta_{(i)}+\theta_{(i)}}{2-\e}\right)
=\cr=-2\beta_\mu C(\e)\frac{\sqrt{\lambda}}{2\pi}\left(\frac{x^\mu_{i}}{\e}+x_{i}^\mu\ln\left(\frac{\mu}{\sqrt{
s_{i,i+1}}}\right)-\frac{\eta_{(i)}+\theta_{(i)}}{2}+O(\e)\right),
\end{gather}
where we have expanded to the significant powers of $\e$. To obtain the total variation we are required to sum
over $i$
\begin{equation}
    \delta_\beta S=-2\beta_\mu
C(\e)\frac{\sqrt{\lambda}}{2\pi}\sum_{i=1}^n\left(\frac{x^\mu_{i}}{\e}+x_{i}^\mu\ln\left(\frac{\mu}{\sqrt{
s_{i,i+1}}}\right)-\frac{\eta_{(i)}+\theta_{(i)}}{2}\right).
\end{equation}
The fact we keep only the significant orders in $\e$ is implicit. The last term clearly cancels in the sum since
the $n$-gon is closed so the sum of displacements of $x^\mu$ vanishes. Hence,
\begin{equation}
    \delta_\beta S=-2\beta_\mu
C(\e)\frac{\sqrt{\lambda}}{2\pi}\sum_{i=1}^n\left(\frac{x^\mu_{i}}{\e}+x_{i}^\mu\ln\left(\frac{\mu}{\sqrt{
s_{i,i+1}}}\right)\right).
\end{equation}
It can be rewritten in the following way,
\begin{gather}\label{final}
    \delta_\beta S=-2\beta_\mu
C(\e)\frac{\sqrt{\lambda}}{2\pi}\times\cr\times\sum_{i=1}^n\left(\frac{x^\mu_{i-1}+x^\mu_{i+1}}{2\e}+
\frac{1}{2}(x^\mu_{i-1}+x^\mu_{i+1})\ln\left(\frac{\mu}{\sqrt{
s_{i,i+1}}}\right)-\frac{1}{2}(x^\mu_{i-1}+x^\mu_{i+1}-2x_{i}^\mu)\ln\left(\frac{\mu}{\sqrt{
s_{i,i+1}}}\right)\right).
\end{gather}
The reason for this rewriting will become clear soon. There are three terms in the brackets of (\ref{final}). To
interpret the first two we note that given two vectors $y^\mu$, $x^\mu$, under a small special conformal
transformation their distance squared transforms as follows
$$(y'-x')^2=(y-x)^2(1-2\beta_\nu(y^\nu+x^\nu))+O(\beta^2).$$
Recall that any $n$ gluon scattering amplitude has a universal divergent part given by a sum over cusps in the
following way \cite{Buchbinder:2007hm}
$$\mathcal{A}_n^{div}=-\sum_{i=1}^n \frac{C(\e)}{\e^2}\frac{\sqrt{\lambda}}{2\pi}\sqrt{\frac{\mu^{2\e}}{(s_{i,i+1})^\e}}.$$
So, some of the terms in (\ref{final}) must be associated to the change from $\mathcal{A}_n^{div}$ to
$\mathcal{A}_n^{'\ div}$ due to the transformation of the invariants $s_{i,i+1}$. This is easily calculated to
$O(\beta^2)$ and we obtain
\begin{gather}\label{divcontrib}
\mathcal{A}_n^{'\ div}=-\sum_{i=1}^n
\frac{C(\e)}{\e^2}\frac{\sqrt{\lambda}}{2\pi}\sqrt{\frac{\mu^{2\e}}{(s_{i,i+1})^\e}}\left(1-2\beta\cdot
(x_{i-1}+x_{i+1})\right)^{-\e/2}=\cr= -\sum_{i=1}^n
\frac{C(\e)}{\e^2}\frac{\sqrt{\lambda}}{2\pi}\sqrt{\frac{\mu^{2\e}}{(s_{i,i+1})^\e}}\left(1+\e\beta\cdot
(x_{i-1}+x_{i+1})\right)=\cr= \mathcal{A}_n^{div}-2\beta C(\e)\frac{\sqrt{\lambda}}{2\pi}\cdot
\sum_{i=1}^n\frac{1}{2\e}\sqrt{\frac{\mu^{2\e}}{(s_{i,i+1})^\e}}(x_{i-1}+x_{i+1})=\cr=
\mathcal{A}_n^{div}-2\beta C(\e)\frac{\sqrt{\lambda}}{2\pi}\cdot
\sum_{i=1}^n\frac{x_{i-1}+x_{i+1}}{2\e}\left(1+\e\ln\left(\frac{\mu}{\sqrt{s_{i,i+1}}}\right)\right),
\end{gather}
where in the last line we have kept only the terms significant in $\e$, again. This manifestly reproduces the
first two terms of (\ref{final}). Ergo, the third term must be associated to a variation of the finite part of
the amplitude.

Denote the finite part of an $n$ point function by $F_n$. Our results prove that it transforms as follows under
a small special conformal transformation
$$\delta_\beta F_n=-\beta_\mu
C(\e)\frac{\sqrt{\lambda}}{2\pi}\sum_{i=1}^n\left((x^\mu_{i-1}+x^\mu_{i+1}-2x_{i}^\mu)\ln\left(\frac{\sqrt{
s_{i,i+1}}}{\mu}\right)\right).$$ Of course, the $\mu$ term cancels and it is natural to get rid of it since
there are no $\mu$s in the finite parts by definition.  In addition, $C=1$ to $O(\e)$ which is all what we have
in the finite parts. Hence, we rewrite it in the following form
\begin{equation}\label{finalfinite part}
    \delta_\beta F_n=-\beta_\mu
\frac{\sqrt{\lambda}}{4\pi}\sum_{i=1}^n(x^\mu_{i-1}+x^\mu_{i+1}-2x_{i}^\mu)\ln\left( s_{i,i+1}\right).
\end{equation}
It can also be presented as a differential equation by substituting
$$\delta_\beta=\beta_\mu\cdot
\sum_i(-2x_i^\mu(x_i\cdot \d_i)+x_i^2\d_i^\mu),$$ we get our final form for the anomalous Ward identity
\begin{equation}\label{diffWard}
    \sum_i\hat{K}_i^\mu F_n=\frac{\sqrt{\lambda}}{4\pi}\sum_{i=1}^n(x^\mu_{i-1}+x^\mu_{i+1}-2x_{i}^\mu)\ln\left( s_{i,i+1}\right).
\end{equation}

Some immediate applications of this result are the fact that four and five point functions are uniquely fixed by
this equation so this establishes the statement in \cite{Alday:2007he} that five point functions at strong
coupling are indeed consistent with the BDS guess \cite{Bern:2005iz}.

We note that combining perturbation theory and strong coupling the equation can be written in general as

\begin{equation}\label{diffWardanycoupling}
    \sum_i(2x_i^\mu(x_i\cdot \d_i)-x_i^2\d_i^\mu)F_n=\frac{f(\lambda)}{4}\sum_{i=1}^n(x^\mu_{i-1}+x^\mu_{i+1}-2x_{i}^\mu)\ln\left( s_{i,i+1}\right),
\end{equation}
where $f(\lambda)$ is the cusp anomalous dimension. In \cite{Drummond:2007wardidentity} the same Ward identity
was proven to hold to all orders of perturbation theory. Here we verify their relation at strong coupling.

Another comment we wish to make is that an analogous exercise to the one we did here can be repeated for the
scaling symmetry. In that case, $r\rightarrow \xi r$ which implies that the area integral is multiplied by an
overall constant $\xi^{-\e}$. This can be easily seen to be completely absorbed by the divergent parts (where it
rescales $s_{i,i+1}\rightarrow \xi^{2}s_{i,i+1}$ for any $i$). Thus, the finite part function satisfies
$$\sum_ix_i\cdot\d_iF_n=0.$$
Similarly, it is annihilated by the Poincar\'e group (since the regulator preserves these symmetries).

\end{document}